\documentclass[journal]{IEEEtran}
\usepackage{amsmath,amssymb}
\usepackage{graphicx}
\usepackage{booktabs}
\usepackage{cite}
\usepackage{url}
\usepackage{microtype}
\usepackage{array}

\begin{document}

\title{RFSS: A Multi-Standard RF Signal Source Separation Dataset
with 3GPP TDL Channels and Hardware Impairments}

\author{Hao~Chen,~Rui~Jin,~and~Dayuan~Tan%
\thanks{H.~Chen, R.~Jin, and D.~Tan are with the Department of Electrical
Engineering. (\textit{Manuscript submitted 2026.})}%
\thanks{Dataset available at
\protect\url{https://huggingface.co/datasets/Chrishao/rfss} (release v1.1).}%
}

\maketitle

% ============================================================
\begin{abstract}
% ============================================================
Cellular standards from 2G to 5G share spectrum, and receivers see several at
once. Source separation could recover the waveforms, but to our knowledge no
public dataset provides per-source references for multi-standard cellular
mixtures. We present
RFSS (RF Signal Source Separation),
100,000 multi-source samples built from the project's own implementations of the
four 3GPP physical-layer specifications (GSM, UMTS, LTE, 5G NR), with
a companion file of 4,000 single-source references. Each sample mixes 2 to 4
sources drawn with replacement, at rates from 2.166 to 122.88 MHz. Every source
passes through its own 3GPP Tapped Delay Line (TDL) multipath fading channel, and
the impairment mode is drawn per source: 19.9 percent of the released sources
receive no impairment, 29.9 percent receive a single impairment type and
50.1 percent receive all six stages, whose two
frequency offsets come from one oscillator error. Sources are
mixed co-channel
or adjacent-channel at a fixed 2 MHz spacing, and the multi-source file is
102.5~GiB in HDF5, split 70/15/15 by index. We benchmark FastICA, Frobenius-norm NMF
and three deep models (Conv-TasNet, DPRNN, an STFT-domain BLSTM) with the
permutation-invariant scale-invariant
signal-to-interference-plus-noise ratio (PI-SI-SINR), and report the gain
over the input mixture. Each deep model uses a validation-chosen learning rate.
On the two-source test split the best deep model improves
PI-SI-SINR by 6.1~dB over the input, where an ideal-ratio-mask oracle reaches
8.0~dB and ICA and NMF reach $-13.6$ and $-1.2$~dB. The dataset, trained
checkpoints and evaluation code are released publicly for non-commercial use.
\end{abstract}


% ============================================================
\section{Introduction}
% ============================================================

Cellular generations share spectrum, and a receiver in a band that operators
reuse across generations faces co-channel and adjacent-channel interference from
signals built to different standards~\cite{andrews2014what}. Spectrum sensing
identifies which
signals are present and of what type. Separation has to go further and
reconstruct each waveform so that a receiver can demodulate it per
standard~\cite{axell2012sensing,goldsmith2009breaking}.

RF source separation runs on both classical and learned methods. Independent
Component Analysis exploits statistical independence between
sources~\cite{hyvarinen2000ica}, and non-negative matrix factorization
factors a magnitude spectrogram into spectral
bases~\cite{fevotte2011nmf}. The Hankel embeddings that this paper's ICA
baseline uses and the STFT frames that its NMF baseline uses are choices of
these baselines and not the only formulations in the literature, and neither
baseline is trained on a corpus. Deep models separate speech with a mask
estimated over long temporal
context, and they are trained with permutation-invariant
losses~\cite{luo2019conv,luo2020dual,kolbaek2017multitalker}. Learned RF
separation is already active: neural separators for co-channel OFDM
mixtures~\cite{lee2023neural}, a time-domain separator that recovers two
unknown radar and communication waveforms~\cite{hinderer2025radar}, a
transformer separator evaluated on the RF Challenge~\cite{lifar2026rf}, and a
large synthetic radio corpus for spectrum sensing~\cite{chang2025csrd}. Training a
separator on multi-standard mixtures still needs per-source references together
with realistic propagation and hardware effects.

Public RF datasets were built for modulation recognition. RadioML provides
single-signal samples over synthetic modulation types and over-the-air
recordings, with no mixtures and no per-source ground
truth~\cite{oshea2016radio,oshea2016convolutional,oshea2018over}. Deep models
classify the signals of individual standards on a single channel or across
distributed sensors~\cite{rajendran2018deep}, and GNU Radio is a toolkit for
building waveform generators that ships no multi-standard mixture
corpus~\cite{blossom2004gnu}.
In audio, WSJ0-2mix, WHAM! and MUSDB18 pair every mixture with its source
signals, which is what made permutation-invariant training
practical~\cite{hershey2016deep,wichern2019wham,rafii2017musdb18}. No public RF
separation dataset covers several cellular standards with per-source
references.

RFSS provides one. It has four features. First, 100,000 multi-source samples
cover GSM, UMTS, LTE and 5G NR, with variation in channel type, source count
and SNR, and every sample stores the per-source waveforms from which the exact
mixture terms can be rebuilt. Second, each standard is generated by the
project's own implementation of the 3GPP physical-layer specification for that
standard instead of as filtered noise or a
generic modulation: GMSK shaping (TS~45.004), Gold-code scrambled Code
Division Multiple Access (CDMA) (TS~25.213), Orthogonal Frequency Division
Multiplexing (OFDM) with the cyclic-prefix durations of TS~36.211
Table~6.12-1, and the flexible numerology of TS~38.211. The GSM symbol rate,
the NR cyclic prefix and the single-user UMTS generator depart from those
specifications, and Section~\ref{sec:construction} names all four departures.
Third, the
hardware impairments use parameter ranges informed by 3GPP base-station
conformance specifications (TS~38.104, TS~36.104, TS~25.104) and are applied
independently to each source, as they would be with several separate
transmitters. Not every source receives the full chain: the impairment mode is
drawn per source, and across the 265,044 released sources 132,902 (50.1 percent)
receive all six stages, 79,314 (29.9 percent) receive a single impairment type
and 52,828 (19.9 percent) receive none. Fourth,
classical and deep methods are evaluated under one protocol, with FastICA,
Frobenius-norm NMF, Conv-TasNet, DPRNN and an STFT-domain BLSTM, an
ideal-ratio-mask oracle as an upper reference, and the learning rate chosen per
model on the validation split. The dataset, generation code, trained
checkpoints and evaluation scripts are released for non-commercial
use.\footnote{Dataset and code:
\url{https://huggingface.co/datasets/Chrishao/rfss}.}

% ============================================================
\section{Related Work}
\label{sec:related}
% ============================================================

Table~\ref{tab:related} compares the public datasets that are used for
separation-style training with RFSS. Every cell is taken from the cited paper
or from this paper.

\begin{table*}[t]
\centering
\caption{Public datasets used for separation-style training, compared with
RFSS. Cells come from the cited papers and from this paper; a dash means the
cited paper does not report the quantity. ``Per-source references'' means the
dataset stores the individual source waveforms of each mixture. The WHAM! size
is the archive size from its distribution page, and the 2,555,904-frame count is
the RadioML 2018.01A release figure, which the cited corpus papers do not state.}
\label{tab:related}
\begin{tabular}{lllrllr}
\toprule
Dataset & Domain & Signals & Mix. & Per-source refs & Src. & Size \\
\midrule
RadioML~\cite{oshea2016radio,oshea2016convolutional,oshea2018over} & RF & 11 or 24 modulations & no & no & 1 & 2,555,904 frames of 1,024 \\
RF Challenge~\cite{lancho2024rfchallenge} & RF & 2 desired, 4 interferers & 8 types & interferer only & 2 & 40,960 samples \\
GNU Radio~\cite{blossom2004gnu} & RF & user-defined & user-built & user-built & user-built & -- \\
WSJ0-2mix~\cite{hershey2016deep,luo2019conv} & Audio & speech & 20,000 train & yes & 2 & 30 h train, 10 h val. \\
WHAM!~\cite{wichern2019wham} & Audio & speech, noise & as WSJ0-2mix & yes & 2 & 17 GB noise set \\
MUSDB18~\cite{rafii2017musdb18} & Music & 4 stems & 150 tracks & yes & 4 & about 10 h, 44.1 kHz \\
\midrule
RFSS (this work) & RF & GSM, UMTS, LTE, 5G NR & 100,000 & yes & 2 to 4 & 102.5 GiB, 2.166-122.88 MHz \\
\bottomrule
\end{tabular}
\end{table*}

The RadioML datasets are the standard corpus for automatic modulation
classification. They contain single transmissions at a range of SNRs, so they
carry no mixture of two signals and no reference for a separator to
reconstruct~\cite{oshea2016radio,oshea2016convolutional,oshea2018over}. The
DARPA Spectrum Collaboration Challenge addressed cooperative spectrum sharing
among radios, which allocates resources rather than recovering waveforms, and
it released no per-source separation labels~\cite{mennes2020sc2}. GNU Radio
is a toolkit for building waveform generators, not a mixture corpus with
per-source references~\cite{blossom2004gnu}. WidebandSig53 has several
overlapping signals per recording from 53 classes, but its labels are classes
and time-frequency boxes, not waveforms~\cite{boegner2022wideband}.

Audio separation research rests on benchmarks that do store the individual
sources. WSJ0-2mix pairs two speakers with per-source signals and motivated
permutation-invariant training~\cite{hershey2016deep,kolbaek2017multitalker}.
WHAM! adds recorded noise~\cite{wichern2019wham}, and MUSDB18 provides four
stems per track~\cite{rafii2017musdb18}. RFSS copies that structure for RF
signals: every mixture carries its per-source waveforms, so the same per-source
structure lets these architectures be trained with their original
loss~\cite{luo2019conv,luo2020dual}. Learned RF separation is now an active
line. Audio-oriented architectures can fail on co-channel OFDM mixtures unless
they are adapted to the OFDM structure~\cite{lee2023neural}, and a
time-domain separator~\cite{hinderer2025radar} and a transformer evaluated on
the RF Challenge~\cite{lifar2026rf} address other RF mixtures. Two large
synthetic corpora exist, but neither is for separation: CSRD2025 (over 25
million frames, 100 modulation types) targets spectrum
sensing~\cite{chang2025csrd}, and RF-GPT trains a language model on
spectrograms with 2 to 5 overlapping signals~\cite{zou2026rfgpt}. None
releases cellular mixtures with per-source waveform references.

The closest RF work is the RF Challenge, a competition at ICASSP~2024 on
interference rejection~\cite{lancho2024rfchallenge}. Each of its mixtures holds
one desired signal of known generative process, either QPSK or OFDM-QPSK, and
one interferer drawn from four signal types, which include recorded
electromagnetic interference and commercial devices. The task is to recover
the desired signal, scored by bit error rate and mean squared error, so the
receiver knows the desired signal type and there is no permutation ambiguity.
RFSS poses the blind
problem instead: every source is an unknown waveform from one of four cellular
standards (with the GSM timing departure of Section~\ref{sec:construction}), the
source count varies from 2 to 4, and the separator resolves the permutation.

To our knowledge, no public dataset addresses blind separation of 2 to 4
simultaneous 3GPP-standard sources at this scale with per-source waveform
references. Section~\ref{sec:benchmark} reports what three transferred deep models
reach on it.

% ============================================================
\section{Dataset Construction}
\label{sec:construction}
% ============================================================

\begin{figure*}[t]
  \centering
  \includegraphics[width=\textwidth]{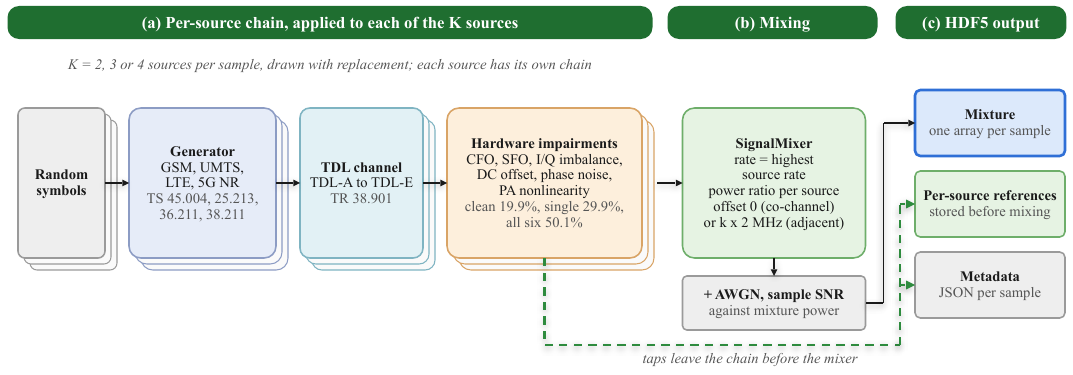}
  \caption{RFSS construction pipeline. Each of the $K$ sources ($K = 2, 3$ or
  $4$, drawn with replacement) passes through signal generation, a 3GPP TDL
  channel and the hardware impairments; the mixer combines them co-channel or
  with adjacent-channel offsets in multiples of 2~MHz, noise is added at the
  sample SNR, and the mixture, per-source references and metadata are written to
  HDF5. References and metadata leave the impairment outputs before mixing.}
  \label{fig:pipeline}
\end{figure*}

Figure~\ref{fig:pipeline} illustrates the construction pipeline. It selects 2--4
source standards at random with replacement, generates a baseband waveform for
each from 3GPP parameters, passes each waveform through an independent TDL channel
and the hardware impairments of its drawn impairment mode, mixes the impaired
signals co-channel or adjacent-channel, and adds noise at the sample SNR. The
stored per-source waveforms are the channel- and impairment-distorted signals at
their native rate, before resampling, frequency shift and power scaling, so the
exact mixture terms used as separation targets are rebuilt from these waveforms
and the metadata (Section~\ref{sec:access}).

\subsection{Signal Generation}

Each waveform is generated at its native sample rate by the project's own
implementation of the 3GPP physical-layer specification for that standard:
GSM at 2.166~MHz,
UMTS at 7.68~MHz, LTE at 1.92--30.72~MHz depending on bandwidth configuration,
and 5G~NR at 15.36--122.88~MHz depending on numerology and bandwidth. Table~\ref{tab:spec} lists the parameters where the implementation departs from the specification.

\begin{table*}[t]
\centering
\caption{Implementation versus specification. ``Specification'' is the value in
the cited 3GPP document, ``implemented'' the value the generator produces. The
consequence column states what the departure changes in the released data.}
\label{tab:spec}
\begin{tabular}{lllll}
\toprule
Standard & Parameter & Specification value & Implemented value & Consequence \\
\midrule
GSM & Symbol rate & 270.833~ksym/s (TS~45.004) & 309.4~ksym/s & Record is 0.873~ms, not 1~ms \\
GSM & Record length & 1~ms (2,166 samples) & 1,890 samples ($270\times7$) & Shortest mixtures in the file \\
UMTS & User count & Multi-user permitted & One user, SF~16, codes 0 & No multi-user case released \\
NR & Cyclic prefix & 176 first, 144 other symbols (TS~38.211) & 143 flat on every symbol & NR record short of 1~ms \\
\bottomrule
\end{tabular}
\end{table*}

\textbf{GSM (2G).} GMSK modulation per 3GPP TS~45.004~\cite{3gpp45004}:
\begin{equation}
\begin{split}
s_{\text{GSM}}(t) &= \exp\!\bigl(j\phi(t)\bigr), \\
\phi(t) &= \pi h \sum_k a_k \int_{-\infty}^{t} g(\tau - kT_s)\,\mathrm{d}\tau
\end{split}
\label{eq:gsm}
\end{equation}
where $h\!=\!0.5$ is the modulation index, $a_k \in \{-1,+1\}$ are
differentially encoded bits, $g(t)$ is the Gaussian pulse shaping filter with
bandwidth-time product $BT\!=\!0.3$, and $T_s\!=\!3.69\,\mu$s is the symbol
duration (270.833~ksps). The exponential form shows that GMSK has a constant
envelope at the generator output: the signal enters only through the cumulative
phase $\phi(t)$. The 200~kHz of TS~45.004 is the carrier spacing of the GSM
channel plan, not the occupied bandwidth: the 99~\% bandwidth of the ideal GMSK
waveform at this symbol rate and $BT$ is 0.247~MHz, which is what the generator
produces before the channel.

The released GSM sources run at 309.4~ksym/s rather than 270.833~ksym/s: the
generator truncates the over-sampling factor to
$\lfloor 2.166\,\text{MHz}/270.833\,\text{ksps}\rfloor = 7$ samples per symbol
instead of 7.998, so a 270-symbol record lasts $270\times7 = 1{,}890$ samples
$= 0.873$~ms (Table~\ref{tab:spec}). All 837 GSM-only mixtures in the multi-source
file and all 1,000 GSM rows of the single-source file carry 1,890 samples. Over
500 released GSM rows the measured
99~\% occupied bandwidth has a median of 0.26~MHz (5th--95th percentile 0.24 to
0.27~MHz), close to the 0.247~MHz of the ideal GMSK spectrum. The channel and the
receiver impairments remove the constant envelope: the per-window
peak-to-average power ratio (PAPR) of the released rows has a median of 0.99~dB
with a 5th--95th percentile spread of 0.06 to 4.38~dB. Measured over the whole
record, the median is 0.93~dB for the 198 clean rows, 1.15~dB for the 79 rows
that draw I/Q imbalance alone and 2.06~dB for the 40 rows that draw DC offset
alone. The multipath channel alone removes the constant envelope: a clean row has
no impairment other than the channel, and it still measures 0.93~dB, so the
fluctuation is not the PA nonlinearity, whose Rapp model leaves a constant
envelope constant.

\textbf{UMTS (3G).} W-CDMA per 3GPP TS~25.211~\cite{3gpp25211}:
\begin{equation}
s_{\text{UMTS}}(t) = \sqrt{P} \sum_k d[k]\, c[k]\, s[k]\; p(t - kT_c)
\label{eq:umts}
\end{equation}
where $P$ is the user power, $d[k]$ are data symbols, $c[k]$ is the Orthogonal
Variable Spreading Factor (OVSF) channelization code, $s[k]$ is the Gold-code
scrambling sequence per TS~25.213~\cite{3gpp25213}, and $p(t)$ is the
root-raised-cosine chip pulse with rolloff $\alpha\!=\!0.22$ and chip duration
$T_c\!=\!260.4$~ns (3.84~Mcps). The generator builds the single-user case: one
QPSK user at spreading factor 16 with OVSF code 0 and scrambling code 0
(Table~\ref{tab:spec}). The 5~MHz of TS~25.104 is the channel spacing; the
released sources occupy a median 4.13~MHz (5th--95th percentile 3.96 to
4.26~MHz), consistent with the 3.84~Mcps chip rate and the 0.22 rolloff of the
RRC pulse.

\textbf{LTE (4G).} OFDM per 3GPP TS~36.211~\cite{3gpp36211}:
\begin{equation}
s_{\text{LTE}}(t) = \sum_{l=0}^{N_{\rm symb}-1}
  \sum_{k=0}^{N_{\rm SC}-1} X_l[k]\,
  e^{j2\pi k\Delta f(t - lT_s - T_{\rm CP})}
\label{eq:lte}
\end{equation}
where $X_l[k]$ are QAM symbols on subcarrier $k$ of OFDM symbol $l$,
$\Delta f\!=\!15$~kHz is subcarrier spacing, and $T_{\rm CP}$ is the cyclic
prefix duration as specified in TS~36.211 Table~6.12-1. The generator uses the
normal-CP lengths of that table: 80 and 72 samples at the 1024-point FFT and 160
and 144 at the 2048-point FFT, so a slot is $80 + 6\times72 + 7\times1024 =
7{,}680$ samples. The dataset draws the channel
bandwidth from 1.4, 3, 5, 10, 15 and 20~MHz (FFT size 1024 at 10~MHz).

\textbf{5G NR (5G).} Flexible-numerology OFDM per 3GPP
TS~38.211~\cite{3gpp38211}:
\begin{equation}
s_{\rm NR}(t) = \sum_{l=0}^{N_{\rm symb}-1}
  \sum_{k=0}^{N_{\rm SC}-1} X_l[k]\,
  e^{j2\pi k\Delta f_{\rm SCS}(t - lT_s^\mu - T_{\rm CP}^\mu)}
\label{eq:nr}
\end{equation}
where $\Delta f_{\rm SCS}\!=\!15\!\times\!2^\mu$~kHz is the subcarrier
spacing for numerology $\mu$. The plotted records use $\mu\!=\!1$ (30~kHz SCS);
the dataset draws $\mu \in \{1,3\}$ (30 or 120~kHz SCS) with bandwidths of 10 to
100~MHz (sample rates up to 122.88~MHz). The cyclic prefix is a flat 7~\% of the
FFT rather than the TS~38.211 value. At the 2048-point FFT of the 50~MHz,
$\mu\!=\!1$ configuration it is $\lfloor 2048\times0.07\rfloor = 143$ samples on
every symbol, where TS~38.211 uses 176 samples on the first symbol of each slot
and 144 on the other thirteen. One 0.5~ms slot of 14 symbols totals 30,720
samples with the specified prefixes and 30,674 implemented, and the released
50~MHz records carry 61,348 samples, two slots, instead of the 61,440 that 1~ms
at 61.44~MHz needs (Table~\ref{tab:spec}). The 30,660 figure of the earlier
audit belongs to the 20~MHz, 1024-point configuration.

\subsection{Channel Modeling}

Each source signal passes through an independent single-input single-output
(SISO) channel drawn from the 3GPP TDL family defined in
TR~38.901~\cite{3gpp38901}, whatever impairment mode the source draws, so the
channel applies to all 265,044 sources: TDL-A 66,279, TDL-B 53,055, TDL-C
39,681, TDL-D 53,105 and TDL-E 52,924. TDL-A and TDL-B are non-line-of-sight delay
profiles. TDL-C is the primary NLOS wideband profile. TDL-D and TDL-E are
Rician (line-of-sight) profiles with K-factors of 13.3~dB and 22~dB
respectively. Channel type is selected from a weighted distribution
per source per sample, with weights
$[0.25, 0.20, 0.15, 0.20, 0.20]$ for TDL-A through TDL-E.

Time-varying Doppler is generated using Jakes' sum-of-sinusoids
model~\cite{jakes1994microwave}, with maximum Doppler frequency drawn from
weighted range categories up to 700~Hz, spanning pedestrian (0--10~Hz),
urban (30--120~Hz), highway (150--300~Hz), and high-speed rail (400--700~Hz)
scenarios. AWGN is added to the mixture after the sources are summed, with SNR
drawn from weighted range categories from $-10$ to 40~dB. The SNR is defined
against the power of the mixture, the mean power of the summed signal handed to
the noise generator, not against any per-source power.

The mixture model combining $N_{\rm src}$ impaired sources is:
\begin{equation}
y(t) = \sum_{i=1}^{N_{\rm src}} \sqrt{P_i}
  \sum_{l=0}^{L_i-1} h_{i,l}(t)\,
  s_i(t - \tau_i - \tau_{i,l})\,
  e^{j2\pi f_i t} + n(t)
\label{eq:mixture}
\end{equation}
where $P_i$ is the power scaling factor for source $i$, $h_{i,l}(t)$ is
the $l$-th tap of its TDL channel impulse response, $\tau_i$ is a
differential timing offset, $f_i$ is the carrier frequency shift
(zero for co-channel, a multiple of 2~MHz for adjacent-channel), and $n(t)$
is additive white Gaussian noise.

\subsection{Hardware Impairments}

Six impairment stages are available for each source before
mixing, informed by 3GPP base-station conformance
requirements~\cite{3gpp38104,3gpp36104,3gpp25104}. The carrier frequency offset
and the sampling frequency offset are derived from one oscillator error, so a
source that receives the full chain carries six stages but five independent
impairment types. The impairment application mode is drawn per source from a
weighted distribution: 20~\% of sources receive no impairment (clean), 30~\% one
type and 50~\% every type. Table~\ref{tab:impairment} gives the realized counts
over all 265,044 sources in the released file. Of the single-impairment sources,
19,837 draw I/Q imbalance, 11,981 DC offset, 11,843 phase noise, 11,825 PA
nonlinearity and 23,828 the combined frequency-offset type, which switches on
the two oscillator-offset stages and leaves the other four at their disabled
defaults. The two offsets switch on and off together, so 156,730 of the 265,044
sources (59.1~\%) carry a nonzero oscillator error.

\begin{table}[t]
\centering
\caption{Impairment modes over the 265,044 released sources. The mode is drawn per
source with weights 0.20, 0.30 and 0.50. A stage is active when its parameter is
off the disabled default.}
\label{tab:impairment}
\begin{tabular}{lrrr}
\toprule
Mode & Share & Count & Active stages \\
\midrule
Clean & 19.9~\% & 52,828 & 0 \\
Single & 29.9~\% & 79,314 & 1 or 2 \\
Multiple & 50.1~\% & 132,902 & 6 \\
\bottomrule
\end{tabular}
\end{table}

\textit{Carrier Frequency Offset (CFO).} A multiplicative phase rotation
$e^{j2\pi f_{\rm CFO} t}$ with $f_{\rm CFO}$ drawn from $\pm0.05$--$5$~ppm
of the carrier, informed by TS~38.104.

\textit{Sampling Frequency Offset (SFO).} Timing drift at the same oscillator
error as the CFO, applied as a resampling of the source by $1 + \varepsilon$
with $\varepsilon$ up to 5~ppm.

\textit{I/Q Imbalance.} Amplitude mismatch 0.1--3~dB and phase error
$1^\circ$--$10^\circ$, informed by the conformance requirements of
TS~36.104.

\textit{Phase Noise.} A Wiener-process phase perturbation with spectral
density $-110$ to $-90$~dBc/Hz at 10~kHz offset, informed by TS~25.104.

\textit{DC Offset.} A constant complex additive offset at $-40$ to
$-30$~dBc relative to signal power, modeling local-oscillator leakage.

\textit{PA Nonlinearity.} The Rapp model~\cite{rapp1991effects} with
input back-off drawn from 3 to 9~dB below the amplifier saturation
point, producing AM/AM and AM/PM distortion.

\subsection{Mixing Scenarios}

Two mixing modes are generated.

\textbf{Co-channel mixing.} All sources are placed at baseband (no frequency
shift: $f_i\!=\!0$ for all $i$). The sources share the whole band, so separation
must rely on differences in modulation type, temporal structure and statistics.
Table~\ref{tab:modes} reports this mode separately from the adjacent-channel one.

\textbf{Adjacent-channel mixing.} Source $k$ of a sample with $N_{\rm src}$ sources is
frequency-shifted by $(k-\lfloor N_{\rm src}/2\rfloor)\times 2$~MHz before summing, the same
spacing for every standard (offsets $\{-2,0\}$ MHz for two sources, $\{-2,0,2\}$ for three,
$\{-4,-2,0,2\}$ for four), and no adjacent-channel leakage model is applied. The spacing is smaller than the bandwidth of most UMTS, LTE and NR
sources, so the sources are displaced but usually still overlap in frequency
(by the nominal channel bandwidths in the metadata, at least one pair of source bands overlaps in 97\% of the two-source, 99.5\% of the three-source and 99.9\% of the four-source adjacent-channel samples).\footnote{Script: \texttt{check/adjacent\_overlap.py}.} The mode is
named for the offset and not for non-overlapping channels.

The number of simultaneous sources $N_{\rm src}$ is drawn from $\{2, 3, 4\}$
with target sampling weights $[0.50, 0.35, 0.15]$. The realized empirical
weights across the 100,000-sample corpus are 0.499, 0.352 and 0.149.
Each source covers about 1~ms at its native rate, except the GSM sources,
which cover 0.873~ms (Section~\ref{sec:construction}). All sources of a sample
are resampled to a common mixture rate equal to the maximum source rate in that
sample, which preserves the full bandwidth of the widest source, and the length
stored in \texttt{signal\_lengths} ranges from 1,890 to 122,880 samples (5G~NR at
122.88~MHz). A source shorter than the mixture is zero-padded at the end of its
stored row: in a 5G~NR and GSM mixture at 61.44~MHz the GSM source covers the
first 53,611 of 61,348 samples, so the last 7,737 samples (0.126~ms) of the
mixture carry no GSM contribution. The 837 mixtures whose sources are all GSM
are 1,890 samples long, the shortest in the file.

% ============================================================
\section{Dataset Characterization}
\label{sec:characterization}
% ============================================================

\subsection{Composition}

The RFSS dataset comprises 100,000 multi-source samples partitioned into three
splits, training (samples 0--69,999, 70~\% of the corpus), validation
(70,000--84,999, 15~\%) and test (85,000--99,999, 15~\%). The validation split
selects the per-model learning rate, and the test split is read once. An
additional 4,000 single-source samples
are provided in a separate file (\texttt{rfss\_single.h5}) for signal
characterization and per-standard classification experiments.

\begin{figure*}[t]
  \centering
  \includegraphics[width=\textwidth]{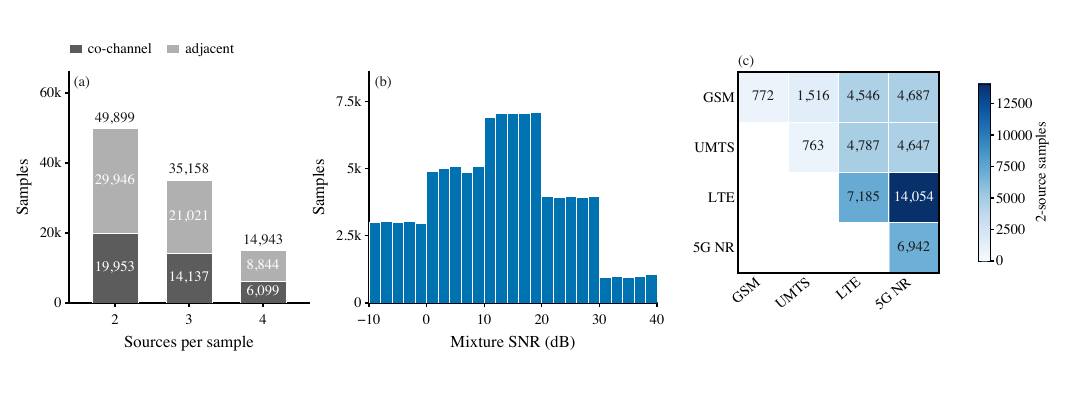}
  \caption{RFSS dataset composition over the 100,000 samples. (a) Sample count
  by source count and mixing mode, counts printed. (b) Mixture SNR distribution
  in 2~dB bins. (c) Standard combination for the 49,899 two-source samples,
  upper triangle including the diagonal. The pairs are unordered, so the
  same-standard pairs (GSM 772, UMTS 763, LTE 7,185, 5G~NR 6,942; 15,662 of
  49,899) appear on the diagonal.}
  \label{fig:dataset_stats}
\end{figure*}

Figure~\ref{fig:dataset_stats} shows the distribution of source counts,
mixing modes, and standard combinations across the corpus. Source count is
weighted toward smaller counts, with realized shares 49.9~\% two-source, 35.2~\%
three-source and 14.9~\% four-source, and mixing modes are weighted toward
adjacent-channel (40~\% co-channel, 60~\% adjacent-channel). For two sources the
standards are drawn with replacement, so the ten unordered pairs are six mixed
pairs and four same-standard pairs (GSM 772, UMTS 763, LTE 7,185 and 5G~NR 6,942
of 49,899 two-source samples).

\subsection{Signal Properties}

\begin{figure*}[t]
  \centering
  \includegraphics[width=\textwidth]{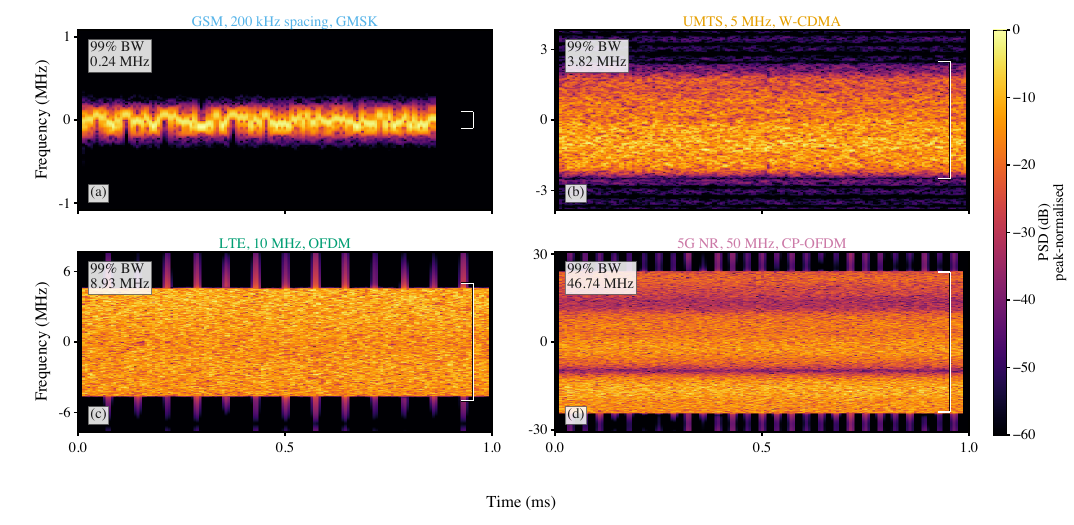}
  \caption{Short-time Fourier transform spectrograms of the four cellular
  standards, each panel on its standard's own native frequency span
  ($\pm f_s/2$: GSM $\pm1.083$, UMTS $\pm3.84$, LTE $\pm7.68$, 5G~NR
  $\pm30.72$~MHz), with its own frequency ticks; shared time axis (0 to 1~ms) and
  one colour scale and bar. Signals are released single-source samples from
  \texttt{rfss\_single.h5} (channel and impairments, before mixing and AWGN) at
  each native rate: GSM 0, UMTS 1000, LTE 2005 (10~MHz), 5G~NR 3003
  ($\mu\!=\!1$, 50~MHz). Each label is the measured 99\,\% bandwidth of that one
  record, not the row median.}
  \label{fig:spectrograms}
\end{figure*}

Figure~\ref{fig:spectrograms} presents representative STFT spectrograms for
each of the four standards, and the distinct time-frequency signatures a
separator must disentangle. GSM appears as a narrow-band trace (0.26~MHz measured
99~\% bandwidth against the 200~kHz carrier spacing of TS~45.004) with
quasi-stationary occupancy. UMTS fills a 4.13~MHz band inside the nominal 5~MHz
channel. LTE and 5G NR show OFDM rectangular spectra, at 15 and 30~kHz
subcarrier spacing in the plotted records.

The largest spectral contrast is between GSM (0.26~MHz measured) and the
wideband OFDM standards, which we expect to give frequency-domain cues that
spectral methods can use; the smallest contrast is between LTE and 5G~NR, which
differ only in subcarrier spacing and bandwidth.

\begin{figure*}[t]
  \centering
  \includegraphics[width=\textwidth]{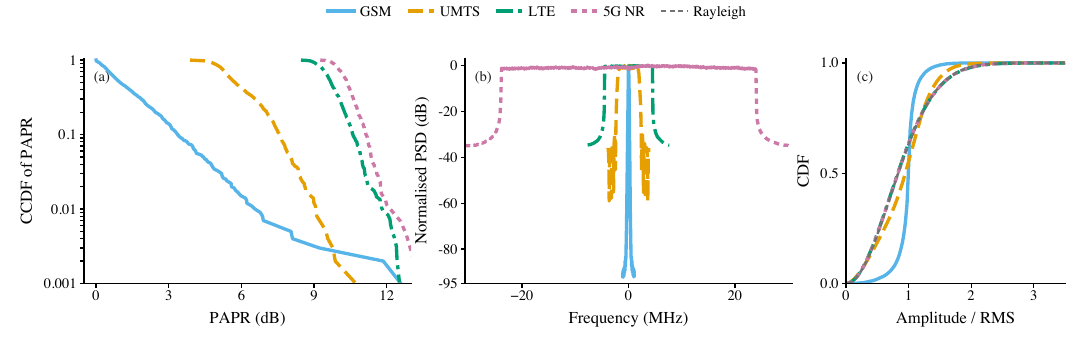}
  \caption{Signal quality of the four cellular standards, from the same
  released samples as Fig.~\ref{fig:spectrograms} (channel and impairments, no
  AWGN) at each native rate. (a) PAPR CCDF pooled over 500 GSM, 334 UMTS, 321 LTE
  and 219 5G~NR rows, over 0.5~ms windows with 50\,\% overlap ($1000$, $1002$,
  $963$ and $438$ windows). (b) Peak-normalised power spectral density,
  Welch-averaged over the same rows. (c) Envelope CDF on the RMS-normalised
  amplitude axis with the Rayleigh reference $1 - \exp(-a^2)$.}
  \label{fig:signal_quality}
\end{figure*}

Figure~\ref{fig:signal_quality} characterizes three signal quality dimensions,
each averaged over a few hundred released rows per standard. The per-window PAPR
over 0.5~ms windows
has median 0.99~dB for GSM (5th--95th percentile 0.06 to 4.38), 5.74~dB for UMTS
(4.65 to 8.00), 9.69~dB for LTE (8.93 to 10.91) and 10.36~dB for 5G~NR (9.66 to
11.35). GSM is not constant-envelope on these rows: the TDL channel, the I/Q
imbalance and the DC offset leave a residual envelope fluctuation of a few dB,
and a clean row, with the channel as its only distortion, still measures 0.93~dB
over the whole record. The power
spectral density panel is computed at each standard's native rate and shows the
99~\% bandwidths, taken over the same rows as medians with 5th--95th
percentiles: 0.26~MHz (0.24 to 0.27) for GSM, 4.13~MHz (3.96 to 4.26) for UMTS,
8.91~MHz (8.61 to 8.97) for LTE and 47.16~MHz (45.17 to 47.58) for 5G~NR. The
nominal channel spacings are 0.2, 5, 10 and 47.9~MHz. The envelope CDF
panel separates GSM (90th percentile at 1.14 times the RMS amplitude) from LTE
and 5G~NR (1.52), which track the Rayleigh reference $1-\exp(-a^2)$.

On the native spans of Fig.~\ref{fig:spectrograms}, the 0.26~MHz GSM band fills
12~\% of the 2.166~MHz axis and the 47.16~MHz NR band fills 76.8~\% of the
61.44~MHz axis. The fractional occupancies differ by a factor of about 6.5, the
absolute bandwidths by a factor of 181, and the mixture rate of each sample is
the maximum of its source rates, up to 122.88~MHz.

\subsection{HDF5 Format and Metadata}

The dataset is stored in two gzip level-6 HDF5 files: the multi-source file
(\texttt{rfss\_dataset.h5}) is 102.5~GiB and the single-source file
(\texttt{rfss\_single.h5}) is 1.3~GiB. The multi-source file contains
four datasets:

\begin{itemize}
  \item \texttt{mixed\_signals}, shape $(100000,\, 122880)$, complex64:
    the observed mixture waveform for each sample.
  \item \texttt{source\_signals}, shape $(100000,\, 4,\, 122880)$,
    complex64: per-source waveforms after the channel and the drawn hardware
    impairments, at their native sample rate, zero-padded in time to 122,880
    samples and to a maximum of four sources. They are not the final mixture
    terms: resampling to the mixture rate, the adjacent-channel frequency shift
    and the power scaling are applied by the released
    \texttt{build\_aligned\_references} function.
  \item \texttt{signal\_lengths}, shape $(100000,)$, int32: the length of
    the resampled mixture signal for each sample, supporting variable-length
    signals within the fixed array.
  \item \texttt{metadata}, shape $(100000,)$, variable-length JSON
    string: per-sample metadata including source count, standard
    identifiers, mixing mode, SNR values, channel type, and hardware
    impairment parameters.
\end{itemize}

The flat array structure (no nested groups) keeps HDF5 indexing simple for
random-access data loading. Complex samples are stored as pairs of float32
(real, imaginary).

% ============================================================
\section{Benchmark Experiments}
\label{sec:benchmark}
% ============================================================

\subsection{Evaluation Metric}
\label{sec:metrics}

We score every method with Permutation-Invariant Scale-Invariant
Signal-to-Interference-plus-Noise Ratio (PI-SI-SINR), following Le Roux et
al.~\cite{leroux2019sdr}. For a reference source $\mathbf{s}$ and an
estimated signal $\hat{\mathbf{s}}$, SI-SINR is:
\begin{equation}
\text{SI-SINR}(\hat{\mathbf{s}},\mathbf{s})
  = 10\log_{10}\frac{\|\alpha\mathbf{s}\|^2}
    {\|\hat{\mathbf{s}} - \alpha\mathbf{s}\|^2},\quad
  \alpha = \frac{\mathrm{Re}\{\hat{\mathbf{s}}^{H}\mathbf{s}\}}{\|\mathbf{s}\|^2}
\label{eq:sisinr}
\end{equation}
where mean subtraction is applied to both signals before the projection to remove
DC bias. The projection coefficient $\alpha$ is a real scalar, so a constant phase
rotation of the estimate lowers the score: the metric is phase sensitive.
Permutation invariance is achieved by taking the permutation of the model outputs
that maximizes the mean SI-SINR over the sources.

The references are the exact per-source terms of the mixture: each stored source
waveform is resampled to the mixture rate, frequency shifted (adjacent-channel
mixtures) and power scaled by the released \texttt{build\_aligned\_references}
function. On 1,000 checked samples the residual between the mixture and the sum of
these references has the power implied by the stored SNR to within 0.2~dB.

We report two quantities. The absolute PI-SI-SINR is the score of the separated
estimates. The gain is the absolute score minus the input score, where the input
score is the mean over the sources of SI-SINR(mixture, source). The mixture contains
every source, so the input score is finite ($-4.4$, $-9.3$ and $-12.0$~dB on average over the
2-, 3- and 4-source test samples). The gain is the main figure because the input level
depends on source count, mixing mode and SNR. For context, Conv-TasNet reaches about
15~dB SI-SNR improvement on WSJ0-2mix~\cite{luo2019conv}.

\subsection{Reference Methods}

\textbf{Oracles.} The noise-limited oracle returns each reference plus the noise
residual (the mixture minus the sum of the references); it is a reference output
that keeps all additive noise in every output. The ideal ratio mask (IRM) oracle
applies masks built from the
reference magnitudes on a 2048-point STFT (hop 512) to the mixture STFT and keeps
the mixture phase. It bounds what time-frequency masking can reach and is the
upper reference of the benchmark.

\textbf{FastICA.} We construct a Hankel (time-delay) embedding matrix
from the scalar observed mixture, increasing the effective number of
observations through lagged copies. The window length is 256 samples
($\approx\!8.3$~$\mu$s at 30.72~MHz) with hop 128 samples, giving a
feature matrix of dimension $n_{\rm frames} \times 512$ (interleaved real
and imaginary channels), with convergence tolerance $10^{-4}$ and up to
500 iterations. The recovered components are matched to the reference sources by
the permutation that maximizes SI-SINR. ICA assumes statistically independent,
non-Gaussian sources, which holds only roughly for cellular signals and becomes
weaker when sources have similar spectra.

\textbf{Frobenius-norm NMF.} The magnitude STFT of the mixture is
factored into $N_{\rm src}$ spectral bases and activations
using NMF with Frobenius norm ($\beta\!=\!2$, 500 iterations, one component per
source). Each source is reconstructed by Wiener-ratio masking in the STFT domain.
NMF can exploit differences in spectral shape, such as GSM's narrow band against
LTE's wide rectangular spectrum. The mask reuses the mixture phase, so NMF quality is
bounded by the spectral diversity of the sources.

\subsection{Deep Learning Methods}

Four architectures, five models counting the Conv-TasNet variant, are trained from scratch on the RFSS training split.

\textbf{Conv-TasNet}~\cite{luo2019conv} uses a learned 1-D convolutional
encoder, a Temporal Convolutional Network (TCN) of dilated depthwise
convolutions that computes the source masks, and a learned decoder. We use
$N\!=\!256$ encoder filters, kernel length $L\!=\!16$, bottleneck dimension
$B\!=\!128$, hidden channels $H\!=\!256$, kernel size $P\!=\!3$, $X\!=\!8$
blocks per repeat and $R\!=\!3$ repeats ($24$ TCN blocks). A sigmoid gives the
masks. A second variant, Conv-TasNet-L256 (a validation-only pilot, seed 0), uses $L\!=\!256$ with stride 64 and is
otherwise identical.

\textbf{DPRNN}~\cite{luo2020dual} splits the encoder output into
overlapping chunks and alternates intra-chunk and inter-chunk
Bidirectional Long Short-Term Memory (BiLSTM) layers. Configuration: $N\!=\!64$
encoder filters, $L\!=\!16$, feature dimension $B\!=\!64$, hidden
dimension $H\!=\!64$, chunk size $P\!=\!50$ encoder frames, 6 dual-path
layers, sigmoid masks.

\textbf{CNN-LSTM} (a validation-only pilot, seed 0) is an encoder-decoder baseline. Three convolutional layers
with channels [64, 128, 256] (kernel size 7, stride 2, batch normalization and ReLU)
extract local features, and a 2-layer BiLSTM with hidden size 256 models the sequence.
A decoder of three transposed convolutions (kernel 8, stride 2) mirrors the encoder,
and a 1$\times$1 convolution estimates each source waveform directly (regression, no
masking). Its first version used a 1$\times$1 convolution with linear interpolation,
whose output is piecewise linear between samples eight apart and which caps the
attainable SI-SINR at 1.98~dB over 300 training crops, so we replaced it.

\textbf{STFT-BLSTM} computes a 2048-point STFT (hop 512) of the complex mixture and
passes the log magnitude through a linear layer and a 2-layer BiLSTM with hidden size
256. It predicts a complex ratio mask per source (tanh on the real and imaginary
parts), applies it to the mixture STFT and inverts the result.

The two-source models have 2.52~M (Conv-TasNet), 2.77~M (Conv-TasNet-L256), 1.11~M (DPRNN),
4.30~M (CNN-LSTM) and 7.36~M (STFT-BLSTM) parameters; the output layers of the 3- and 4-source models are larger (Table~\ref{tab:cost}).

All models are trained with permutation-invariant training
(PIT)~\cite{kolbaek2017multitalker}, minimizing the negative SI-SINR of the best
permutation. Optimization uses Adam, gradient clipping at norm 1.0, batch size 8 and
cosine annealing to $\eta_{\min}\!=\!10^{-5}$ over $E\!=\!10$ epochs. Input crops of
7,680 samples are redrawn at random every epoch. The learning rate is chosen per
model on the validation split (Appendix~\ref{app:training}) and is not shared across models. In an earlier run at $10^{-3}$ the Conv-TasNet did not leave a plateau within 30 epochs. In
a validation sweep with otherwise identical settings it stayed at $+2.88$ and
$+2.89$~dB over the first two epochs at $10^{-3}$ and left that plateau in the first epoch
at $3\times10^{-4}$ (run \texttt{l16} in \texttt{check/encoder\_sweep\_results.json},
fields \texttt{epochs[0..1].val.all.gain\_over\_input\_db}). For the STFT-BLSTM at $10^{-3}$, whether the model left the plateau
depended on the seed. We therefore choose the learning rate with the higher worst-seed validation
gain at the common epoch budget. The checkpoint with the lowest validation loss is
evaluated once on the test split. Each model trains for $E$ passes over the training subset
with its source count, so the 3-source and 4-source models see 0.70 and 0.30 times the
optimizer steps of the 2-source model (about 4,364, 3,069 and 1,318 steps per epoch for the
34,912, 24,547 and 10,541 training samples). All experiments use single-channel (SISO)
signals.

\textbf{Flagged cells.} The 3-source and 4-source cells were checked against the 2-source curve of the same model: at the same
number of optimizer steps, a validation gain more than $0.5$~dB below the (interpolated) 2-source gain, or a rise of less than $0.3$~dB since epoch 2,
flagged the cell and triggered a validation-only screen of neighbouring learning rates. The 3- and 4-source tasks are harder, so the reference is a diagnostic and not a target, and
no cell was dropped for being low. Five cells were flagged; the per-cell record is in Appendix~\ref{app:training}. One screen decided a learning rate: the DPRNN 3-source run at $10^{-3}$ stayed at $+3.05$~dB for four epochs, and a screen of four rates favoured $5\times10^{-4}$ over $10^{-3}$ by $0.51$~dB at epoch 2
(borderline by the preset rule of $0.5$~dB, supported by the monotone trend with the learning rate and by the stuck run), so the cell was restarted at $5\times10^{-4}$ and ended at $+4.92$~dB.

\subsection{Evaluation Protocol}

Every method is scored on one random 7,680-sample window of each test signal (shorter
signals in full), chosen by a fixed seed (crop seed 0) and the sample index, which is the
crop definition of the validation protocol. The same 7,526, 5,324 and 2,150 test samples are used for
the 2-, 3- and 4-source columns, against the
references described above. Intervals are 95\% bootstrap intervals over samples, and
differences between methods are paired bootstraps over the same samples. Three further passes
(the first 7,680 samples of each signal, and random windows with seeds 1 and 2) change the gain of
no 2-source model by more than 0.06~dB and of no 3- or 4-source model by more than 0.09~dB, and leave the order of the
models unchanged in every source count; ICA and NMF were scored in the first two passes only. The
first-window pass was run and read before the crop protocol was fixed; the headline is crop seed 0, and all four passes are
reported. All recipe decisions (learning rate, epoch budget, checkpoint) use the
validation split only. Two interim evaluations were run on the test split before the
final run: one on 600 two-source samples with an early checkpoint, and one of ICA and NMF on
the full test split. Neither was used to choose any setting.
Before the final runs we fixed one target for the training budget: a validation gain of at least $+6$~dB in the adjacent-channel, SNR${>}20$~dB bin after two epochs. No model's estimated gain reached it (best: STFT-BLSTM, $+5.68$~dB, 95\% interval $[+4.65,+6.73]$, which includes $+6$~dB). Exploratory runs of 10 epochs exceeded $+6$~dB, so the epoch budget was set from the validation curves, which departs from the pre-registered plan. We chose $E\!=\!10$ by a threshold fixed in advance: a 20-epoch probe of STFT-BLSTM on the validation split had to gain at least $0.30$~dB over the 10-epoch run in the all-samples bin. It gained $0.20$~dB (95\% interval $[0.15,0.25]$). Within the probe, the second 10 epochs added $0.26$~dB ($[0.20,0.31]$), so 10 epochs under-train the models relative to 20; the threshold is a cost--benefit rule, not a convergence claim.

\subsection{Results}

Table~\ref{tab:main} reports the gain over the input mixture by source count.

\begin{table}[t]
\centering
\caption{Gain over the input mixture (dB, PI-SI-SINR) on the test split, by source
count (random 7,680-sample windows, crop seed 0). Deep models at 2 sources: mean over three
seeds (range of the seeds); deep models at 3 and 4 sources: one seed, 95\% bootstrap interval;
all other rows: 95\% bootstrap interval over test samples. The input level is $-4.4$ / $-9.3$ / $-12.0$~dB.}
\label{tab:main}
\resizebox{\columnwidth}{!}{%
\begin{tabular}{lrrr}
\toprule
Method & 2-source & 3-source & 4-source \\
\midrule
ICA & $-$13.60 [$-$13.70, $-$13.50] & $-$11.09 [$-$11.20, $-$10.97] & $-$9.49 [$-$9.67, $-$9.31] \\
NMF & $-$1.24 [$-$1.31, $-$1.19] & $-$0.57 [$-$0.63, $-$0.51] & $-$0.05 [$-$0.14, +0.05] \\
Conv-TasNet & +5.61 (+5.42 to +5.72) & +4.77 [+4.68, +4.86] & +3.54 [+3.45, +3.63] \\
DPRNN & +5.85 (+5.76 to +5.90) & +5.03 [+4.93, +5.13] & +3.44 [+3.35, +3.52] \\
STFT-BLSTM & +6.14 (+6.12 to +6.15) & +5.14 [+5.04, +5.24] & +4.52 [+4.40, +4.63] \\
\midrule
IRM oracle & +8.04 [+7.92, +8.16] & +9.61 [+9.48, +9.75] & +10.35 [+10.13, +10.55] \\
Noise-limited oracle & +10.16 [+9.98, +10.35] & +11.53 [+11.29, +11.79] & +12.13 [+11.72, +12.52] \\
\bottomrule
\end{tabular}
}
\end{table}

At two sources the STFT-BLSTM gains $+6.14$~dB (mean of three seeds), which is 76\% of the
$+8.04$~dB of the IRM oracle; DPRNN and Conv-TasNet reach $+5.85$ and $+5.61$~dB. ICA and NMF end below the
input level ($-13.60$ and $-1.24$~dB). In every matched seed pair (same seed number, independently
initialised models) the STFT-BLSTM exceeds DPRNN, by $+0.29$~dB on average (range $+0.25$ to $+0.37$), and DPRNN exceeds Conv-TasNet, by $+0.24$~dB
(range $+0.17$ to $+0.33$); the paired 95\% intervals over test samples are about $\pm0.03$~dB wide, and the sign is the same in the
adjacent-channel and co-channel subsets with SNR above 20~dB. We therefore state the order STFT-BLSTM $>$ DPRNN $>$ Conv-TasNet for this recipe, with two cautions: the
seed pairing is nominal, so the evidence is the consistent sign in three pairs together with the small seed ranges ($0.03$, $0.15$ and $0.29$~dB), and the models differ in
parameters ($7.36$~M, $1.11$~M, $2.52$~M) and in cost per epoch, so the comparison does not isolate the domain in which the mask is estimated.
With three and four sources every model gains less, and the gap to the IRM oracle, which rises to $+9.61$ and $+10.35$~dB, grows. These cells have one seed each, so we
give no order and no trend across source counts. At three sources the models lie within $0.37$~dB (STFT-BLSTM minus DPRNN $+0.11$~dB, $[+0.06,+0.15]$);
at four sources the STFT-BLSTM leads the two time-domain models by about $1$~dB ($+1.08$~dB, $[+1.01,+1.15]$, over DPRNN), while DPRNN and Conv-TasNet differ by $-0.10$~dB ($[-0.14,-0.06]$).

Table~\ref{tab:modes} splits the 2-source gain by mixing mode and shows the
adjacent-channel, SNR above 20~dB bin in which the epoch-budget criterion was set.

\begin{table}[t]
\centering
\caption{Two-source gain over the input (dB) by mixing mode on the test split.
$N$ is the number of test samples in each column.}
\label{tab:modes}
\resizebox{\columnwidth}{!}{%
\begin{tabular}{lrrrr}
\toprule
Method & All & Co-channel & Adjacent & Adjacent, SNR${>}20$~dB \\
\midrule
ICA & $-$13.60 & $-$13.54 & $-$13.65 & $-$16.01 \\
NMF & $-$1.24 & $-$1.11 & $-$1.34 & $-$2.61 \\
Conv-TasNet & +5.61 & +5.68 & +5.56 & +6.21 \\
DPRNN & +5.85 & +5.89 & +5.82 & +6.58 \\
STFT-BLSTM & +6.14 & +6.13 & +6.14 & +6.94 \\
\midrule
IRM oracle & +8.04 & +7.77 & +8.22 & +10.86 \\
$N$ & 7{,}526 & 3{,}035 & 4{,}491 & 1{,}105 \\
\bottomrule
\end{tabular}
}
\end{table}

Averaged over all SNRs the deep models gain about the same on co-channel and on adjacent-channel mixtures
(STFT-BLSTM $+6.13$ and $+6.14$~dB, DPRNN $+5.89$ and $+5.82$~dB, Conv-TasNet $+5.68$ and $+5.56$~dB), whereas the IRM oracle gains $+7.77$ and
$+8.22$~dB; adjacent-channel mixtures are not easier for these models, although they are for the oracle. In the adjacent-channel, SNR${>}20$~dB bin the
three models reach $+6.94$, $+6.58$ and $+6.21$~dB against $+10.86$~dB for the oracle. Each of the nine two-source runs gains at least $+5.42$~dB over all test samples and at least
$+5.94$~dB in that bin. The pre-registered criterion concerned two-epoch runs on the validation split, so these test values neither meet nor fail it.

The learning-rate screening and the two validation-only variants are recorded in Appendix~\ref{app:training}.

Table~\ref{tab:cost} gives the training cost. Epoch times are raw wall-clock values with two to three training runs sharing the machine; they are not
single-job timings.

\begin{table}[t]
\centering
\caption{Training cost per epoch (minimum / median / maximum over the epochs of the final runs; two-source: seeds 1 and 2,
otherwise seed 0) and the number of other final training runs at the same time.
The maximum of the 2-source Conv-TasNet includes two manual pauses of one run.
Scoring jobs, screens and the seed-0 pilots are not counted. STFT-BLSTM ran on the
CPU, the others on the GPU of an Apple M4 Pro with 48~GB memory.}
\label{tab:cost}
\resizebox{\columnwidth}{!}{%
\begin{tabular}{llrrrr}
\toprule
Model & Sources & Parameters & Device & Epoch time (s) & Other runs \\
\midrule
Conv-TasNet & 2 & 2.52~M & MPS & 1{,}424 / 1{,}676 / 6{,}874 & 1--3 \\
Conv-TasNet & 3 & 2.56~M & MPS & 1{,}134 / 1{,}385 / 1{,}511 & 1--3 \\
Conv-TasNet & 4 & 2.59~M & MPS & 610 / 695 / 780 & 2--3 \\
DPRNN & 2 & 1.11~M & MPS & 672 / 823 / 1{,}063 & 1--3 \\
DPRNN & 3 & 1.12~M & MPS & 626 / 768 / 794 & 0--2 \\
DPRNN & 4 & 1.12~M & MPS & 378 / 421 / 436 & 2--2 \\
STFT-BLSTM & 2 & 7.36~M & CPU & 502 / 537 / 566 & 0--3 \\
STFT-BLSTM & 3 & 9.46~M & CPU & 522 / 534 / 604 & 3--3 \\
STFT-BLSTM & 4 & 11.56~M & CPU & 284 / 292 / 293 & 3--3 \\
\bottomrule
\end{tabular}
}
\end{table}

\subsection{SNR Sensitivity Analysis}
\label{sec:snr}

Table~\ref{tab:snr} stratifies the 2-source gain by mixture SNR.

\begin{table}[t]
\centering
\caption{Two-source gain over the input (dB) by mixture SNR on the test split.}
\label{tab:snr}
\resizebox{\columnwidth}{!}{%
\begin{tabular}{lrrrr}
\toprule
Method & $-10$--$10$~dB & 10--20~dB & 20--30~dB & ${>}30$~dB \\
\midrule
ICA & $-$10.99 & $-$14.79 & $-$15.91 & $-$16.81 \\
NMF & $-$0.16 & $-$1.58 & $-$2.38 & $-$3.08 \\
Conv-TasNet & +4.98 & +5.88 & +6.11 & +6.73 \\
DPRNN & +5.16 & +6.13 & +6.41 & +7.11 \\
STFT-BLSTM & +5.43 & +6.37 & +6.75 & +7.66 \\
\midrule
IRM oracle & +5.69 & +8.92 & +10.27 & +11.71 \\
\bottomrule
\end{tabular}
}
\end{table}

All three deep models improve with SNR, the STFT-BLSTM from $+5.43$~dB in the lowest bin to $+7.66$~dB above 30~dB, DPRNN from $+5.16$ to $+7.11$~dB and
Conv-TasNet from $+4.98$ to $+6.73$~dB. The IRM oracle rises faster, from $+5.69$ to $+11.71$~dB, so the deficit of the STFT-BLSTM grows from $0.3$ to
about $4$~dB; in the lowest bin the three models are within $0.8$~dB of the oracle. ICA and NMF fall with SNR ($-10.99$ to $-16.81$~dB and $-0.16$ to $-3.08$~dB); we did not
isolate why.

% ============================================================
\section{Data Access and Format}
\label{sec:access}
% ============================================================

\subsection{Public Release}

Both HDF5 files (\texttt{rfss\_dataset.h5}, 102.5~GiB; \texttt{rfss\_single.h5},
1.3~GiB) are hosted on Hugging Face at
\url{https://huggingface.co/datasets/Chrishao/rfss} under release tag v1.1: data
under CC BY-NC 4.0, code under PolyForm Noncommercial 1.0.0. The 15 trained
checkpoints scored in this paper are in the \texttt{checkpoints} folder of that
release, and the generation and evaluation code is at
\url{https://github.com/chenhao1umbc/dataset_RFSS}.

\subsection{Relation to Earlier arXiv Versions}

This paper supersedes arXiv:2508.12106 and arXiv:2604.00398 by the same group.
The first reported separation results (26.7~dB on 52,847 samples) that could not
be reproduced from the released data and code; the results here replace them.
The second described the dataset with errors (common sampling rate, signal
duration, channel and Doppler ranges, SNR range, mixing-mode shares and GSM
peak-to-average power ratio), which Sections~\ref{sec:construction}
and~\ref{sec:characterization} correct. Release v1.1 holds the same HDF5 files as
v1.0, with the corrected documentation, the checkpoints and the licences.

\subsection{HDF5 Access Example}

One call loads a sample from either file; the released
\texttt{rfss\_single.h5} is used here because it is the smaller download.

\begin{verbatim}
import h5py, json
with h5py.File('rfss_single.h5', 'r') as f:
    x = f['mixed_signals'][idx]
    s = f['source_signals'][idx]
    n = f['signal_lengths'][idx]
    m = json.loads(f['metadata'][idx])
\end{verbatim}

The train/validation/test split is applied in code by index range:
training uses indices 0--69,999, validation 70,000--84,999, and test
85,000--99,999. No separate split files are needed.

\subsection{Evaluation Code}

A \texttt{SeparationDataset} class is provided with evaluation scripts for
the classical baselines and the deep models; the scripts retrain the models and
reproduce the tables in Section~\ref{sec:benchmark} from the released dataset. A
requirements file lists the Python dependencies, including PyTorch, h5py,
scikit-learn and scipy.

% ============================================================
\section{Limitations and Future Work}
\label{sec:limitations}
% ============================================================

\textbf{Reference construction.} The released arrays store each source after the channel and
hardware impairments but before resampling, frequency shift and power scaling, so a
user must rebuild the exact mixture terms with
\texttt{build\_aligned\_references}; storing the final terms as well would remove
that step in a later release.

\textbf{Synthetic channel.} Every source passes through a synthetic single-input
single-output (SISO) TDL channel from TR~38.901 with Jakes Doppler: statistical
profiles, not measurements, with no site-specific geometry, spatial correlation or
line-of-sight steering. A geometry-based model is left to a later release.

\textbf{Impairments, UMTS user count and the GSM/NR departures.} The impairment
mode is drawn per source, so 19.9~\% of the released sources are clean and
29.9~\% carry a single impairment type; only the 50.1~\% with mode ``multiple''
receive the full six-stage chain (Table~\ref{tab:impairment}). The UMTS generator
builds a single user at spreading factor 16 with OVSF code 0 and scrambling code
0, so no multi-user UMTS case is released. The GSM generator truncates the
over-sampling factor to an integer, so the released GSM sources run at
309.4~ksym/s instead of the 270.833~ksym/s of TS~45.004 and last 0.873~ms instead
of 1~ms; the NR cyclic prefix is a flat 7~\% of the FFT, 143 samples at the
2048-point FFT against 176 and 144 in TS~38.211, so the NR records are short of
1~ms. Table~\ref{tab:spec} lists all four departures; LTE uses the TS~36.211
normal-CP lengths exactly.

\textbf{Evaluation scope.} All methods are scored on one 7,680-sample window of
each signal, and the metric is phase sensitive: the projection coefficient is
real, so a constant phase rotation of the estimate lowers the score
(\texttt{src/baseline\_algorithms.py:50}). Comparisons rest on three seeds per
model (two sources) or one seed (three and four sources) from one learning-rate
grid per model; seed 0 of the two-source runs also chose the learning rates, so
its score is mildly optimistic. No model is shown to have converged, the cosine
schedule anneals to almost zero, and the epoch budget was set from the validation
curves after the pre-registered criterion was missed; a 20-epoch probe gained
0.20~dB in the all-samples bin over 10 epochs, so some performance was left
unused. The 3- and 4-source models take fewer optimizer steps than the 2-source
models because each trains for the same number of epochs on a smaller subset.

\textbf{Adjacent-channel mode.} The mode shifts the sources by multiples of 2~MHz for
every standard and applies no filtering or adjacent-channel leakage model: a
frequency displacement, not a guard-banded channel plan. The 2~MHz spacing is fixed
with no adjacent-channel interference ratio (ACIR), and 97\% of the two-source
adjacent-channel mixtures (99.5\% of the three-source and 99.9\% of the four-source
ones) have overlapping nominal channel bands. For 1.8\% of them the mixture rate is
below 4~MHz (GSM at 2.166~MHz and narrow LTE channels at 1.92 and 3.84~MHz), so the
2~MHz shift wraps around the sampling band; the references use the same shift, so
the labels are exact, but these mixtures are effectively co-channel or folded
scenes.

\textbf{Single-antenna (SISO) only.} The dataset models each source as a scalar
complex waveform with no spatial structure, so the metadata's MIMO field is
recorded but unused. A $2\times2$ or $4\times4$ extension with spatial correlation
per 3GPP TDL would allow spatial separation methods.

\textbf{Waveform coverage.} The dataset covers GSM, UMTS, LTE and 5G~NR;
NB-IoT, LTE-M, sidelink and ISAC waveforms are absent, and so are source counts
above four and bandwidths above 100~MHz.

% ============================================================
\section{Conclusion}
\label{sec:conclusion}
% ============================================================

We presented RFSS, 100,000 multi-source RF samples that cover GSM, UMTS, LTE and
5G NR. Each source is generated by the project's own implementation of the 3GPP
physical-layer specification for its standard, passed through its own TDL
fading channel and a hardware impairment chain whose carrier and
sampling frequency offsets come from one oscillator error, and mixed co-channel
or adjacent-channel at a fixed 2~MHz spacing. The impairment mode is drawn per
source, so 132,902 of the 265,044 released sources (50.1~\%) carry the full
six-stage chain, 79,314 (29.9~\%) carry one type and 52,828 (19.9~\%) are clean.
Every sample stores the per-source waveforms from which the exact mixture terms are
rebuilt, so a model can be trained with permutation-invariant supervision.

On the two-source test split the best deep model, the STFT-BLSTM, improves PI-SI-SINR by $+6.14$~dB
over the input mixture, against $+8.04$~dB for the ideal-ratio-mask oracle
and $-13.60$ and $-1.24$~dB for ICA and NMF. The three deep models keep the same order in every matched seed pair, with mean
paired differences of $0.24$ to $0.53$~dB; with three and four sources the gains fall to $+5.14$ and $+4.52$~dB for the best model
(one seed per cell). Training was sensitive to the
learning rate and, for some models, to the seed; the learning-rate screening is in
Appendix~\ref{app:training}. The dataset, the trained checkpoints and the
evaluation code are released for non-commercial use at
\url{https://huggingface.co/datasets/Chrishao/rfss} (release v1.1).

\appendix

% ============================================================
\section{Training Diagnostics}
\label{app:training}
% ============================================================

Table~\ref{tab:lr} gives the learning-rate screening per model, and
Table~\ref{tab:flagged} the per-cell record of the flagged-cell rule. Two further
variants were trained for 10 epochs with seed 0 and scored on the 800 validation
crops only: Conv-TasNet-L256 ($+4.51$~dB, 95\% interval $[+4.16,+4.86]$) and the
CNN-LSTM with the transposed decoder ($+4.47$~dB, $[+4.11,+4.81]$), both below the
three test-table models on the same crops ($+5.84$, $+5.99$ and $+6.36$~dB for
Conv-TasNet, DPRNN and STFT-BLSTM, seed 0).

\begin{table}[t]
\centering
\caption{Learning-rate screening on the validation split (800 fixed crops,
2-source): the adjacent-channel, SNR${>}20$~dB gain (dB) at the screening epoch
and ``escape'', the first epoch above 4~dB. The chosen rate is in bold; gains are
the lowest over the seeds at the second epoch (``--'': none).}
\label{tab:lr}
\resizebox{\columnwidth}{!}{%
\begin{tabular}{lrrr}
\toprule
Model & Learning rate & Gain & Escape epoch \\
\midrule
Conv-TasNet & $10^{-4}$, $\mathbf{3\times10^{-4}}$, $10^{-3}$ & +4.47, \textbf{+4.84}, +2.40 & 2, \textbf{1}, -- \\
Conv-TasNet-L256 & $3\times10^{-5}$, $\mathbf{10^{-4}}$, $3\times10^{-4}$ & +0.89, \textbf{+2.50}, +1.55 & --, \textbf{--}, -- \\
DPRNN & $10^{-4}$, $3\times10^{-4}$, $\mathbf{10^{-3}}$ & +4.47, +4.93, \textbf{+5.09} & 2, 1, \textbf{1} \\
CNN-LSTM (transposed decoder) & $10^{-4}$, $\mathbf{3\times10^{-4}}$, $10^{-3}$ & +2.14, \textbf{+2.77}, +2.44 & --, \textbf{--}, -- \\
STFT-BLSTM & $\mathbf{3\times10^{-4}}$, $10^{-3}$ & \textbf{+5.53}, +2.46 & \textbf{1}, -- \\
\bottomrule
\end{tabular}
}
\end{table}

\begin{table}[t]
\centering
\caption{Flagged-cell record: every 3- or 4-source cell the rule marked, its screen
outcome and its final validation gain (all-samples bin, dB).}
\label{tab:flagged}
\resizebox{\columnwidth}{!}{%
\begin{tabular}{llllr}
\toprule
Model & Src. & Flagged at & Screen outcome & Final gain \\
\midrule
STFT-BLSTM & 3 & 2 & cleared at epoch 2 & +5.02 \\
Conv-TasNet & 3 & 2, 4 & slow, not stuck (screen: none) & +4.62 \\
Conv-TasNet & 4 & 4 & three rates, no change & +3.47 \\
DPRNN & 3 & 2, 4 & restarted at $5\times10^{-4}$ & +4.92 \\
DPRNN & 4 & 4 & three rates, no winner & +3.41 \\
\bottomrule
\end{tabular}
}
\end{table}

\bibliographystyle{ieeetr}
\bibliography{main}
\end{document}